\documentclass[
preprint,
amsmath,amssymb,
prb,
]{revtex4-2}

\usepackage{graphicx}% Include figure files
\usepackage{dcolumn}% Align table columns on decimal point
\usepackage{bm}% bold math
\begin{document}

%\preprint{APS/123-QED}

%\title{Electrostatically controlled band engineering in graphene bipolar waveguides and their THz applications}% Force line breaks with \\
\title{Bipolar electron waveguides in graphene}

\author{R. R. Hartmann}
\email{richard.hartmann@dlsu.edu.ph}
\affiliation{
Physics Department, De La Salle University,  2401 Taft Avenue, 0922 Manila, Philippines
}

\author{M. E. Portnoi}
\email{M.E.Portnoi@exeter.ac.uk}
\affiliation{Physics and Astronomy, University of Exeter, Stocker Road, Exeter EX4 4QL, United Kingdom}
\affiliation{ITMO University, St. Petersburg 197101, Russia}


\begin{abstract}
%We show via a simple analytic theory, supported by numerical experiment, 
We show analytically that the ability of Dirac materials to localize an electron in both a barrier and a well can be utilized to open a pseudo-gap in graphene's spectrum. By using narrow top-gates as guiding potentials, we demonstrate that graphene bipolar waveguides can create a non-monotonous one-dimensional dispersion along the electron waveguide, whose electrostatically controllable pseudo-band-gap is associated with strong terahertz transitions in a narrow frequency range.
\end{abstract}

\maketitle

\section{Introduction}
Graphene's gapless, relativistic spectrum leads to many unusual transport properties~\cite{castro2009electronic}, such as Klein tunneling~\cite{katsnelson2006chiral} and the suppression of back scattering~\cite{ando1998berry}. Graphene also exhibits strong optical transitions, with a universal absorption of $\pi e^2/ \hbar c \approx 2.3 \%$ over a broad range of frequencies~\cite{nair2008fine}. These defining features of graphene make electronic and optical control difficult to achieve. Indeed, for the realisation of optoelectronic devices which capitalize on the relative nature of graphene's dispersion, one must overcome certain undesirable relativistic features, without destroying all the attractive effects.

%For optoelectronic devices applications, it would be highly  desirable to modify graphene's spectrum such that it possesses valence-like and conduction-like bands, separated by an externally tunable pseudo-bandgap (a true gap would require a drastic modification of the material by either functionalization, cutting or rolling), as well as reduce the freedom of quasiparticle  motion from two-dimensions down to one. 
For optoelectronic devices applications, it would be highly  desirable to modify graphene's spectrum such that it possesses valence-like and conduction-like bands, separated by an externally tunable pseudo-bandgap (a true gap would require a drastic modification of the material by either functionalization, cutting or rolling), as well as reduce the freedom of quasiparticle  motion from two-dimensions down to one. In essence, we wish to form an analog of a narrow-gap carbon nanotube spectrum, without physically deforming the sheet by cutting it to make a ribbon, or rolling it to form a nanotube. In this paper, we propose a setup based on a bipolar potential, which does not open a true bandgap. However, at certain parameters of the potential, the electron dispersion becomes non-monotonous, and has pronounced extrema associated with avoided crossings between the states confined within a barrier and a well. We call the energy separation at a local dispersion minimum a pseudo-gap. Such a pseudo-gap would result in the giant enhancement of the probability of optical transitions~\cite{hartmann2019interband}. However, unlike a nanotube or ribbon, whose bandgap is predefined by geometry, we seek to create a pseudo-gap which is fully tunable, without the need for huge magnetic fields~\cite{portnoi2008terahertz,hartmann2015terahertz}.  
Rather than achieving the quantization of momentum through geometry like a nanoribbon, or nanotube does, one may instead quantize momentum via the application a quasi-one-dimensional (1D) electrostatically defined potential, i.e., by using electron waveguides~\cite{pereira2006confined,cheianov2007focusing,tudorovskiy2007spatially,shytov2008klein,beenakker2009quantum,zhang2009guided,hartmann2010smooth,zhao2010proposal,sharma2011electron,hartmann2014quasi,he2014guided,hasegawa2014bound,xu2015guided,xu2016guided,mondal2019thz}. Unlike a physical tube whose radius cannot be changed, externally applied potentials can be easily varied. 
%One way to reduce the freedom of graphene's quasiparticle  motion is to subject them to 
%a quasi-one-dimensional (1D) electrostatically define potential, i.e., an electron waveguide.
%Indeed, t
There has been significant experimental progress since the pioneering work in the field of graphene electron waveguides~\cite{huard2007transport, ozyilmaz2007electronic, gorbachev2008conductance,liu2008fabrication,williams2011gate,rickhaus2015guiding} and the recent breakthrough of utilizing a nanotube as a top gates~\cite{cheng2019guiding} enabled the detection of individual guided modes within a single waveguide. However, apart from graphene waveguides possessing a threshold in the characteristic potential strength required to observe a fully bound mode~\cite{hartmann2010smooth,hartmann2017two}, one could argue that a single graphene electron waveguides provides similar physics to that studied in quasi-1D channels within conventional semiconductor systems. Indeed, the absolute value of electron momentum along the direction of a waveguide formed by an attractive potential (quantum well) defines the electron's effective mass. Confined states of a deep well start with negative energy and for large values of momentum have a positive dispersion along the waveguide. Similarly, a potential barrier can also form a waveguide, where the confined electron states have negative dispersion, i.e., are hole-like. This raises the question what happens when branches of negative and positive dispersion meet? Answering this question is the focus of this paper. 
In what follows we show that a bipolar electronic waveguide is a fully tunable quasi-1D system with a non-monotonous dispersion accompanied by pseudo-gaps, characterized by a giant enhancement of density of states and interband dipole transition probabilities in the energy range where graphene's own density of states is rather small. 
In addition we present a general analytic formalism allowing us to find with a spectacular degree of accuracy the main features of a bipolar waveguide from the properties of a single quantum well.

%In this Letter we propose to use bipolar electrostatically-induced waveguides as a fully tunable quasi-1D system with a non-monotonous dispersion characterized by a giant enhancement of density of states in the energy range where graphene's own density of state is rather small.

%the two branches mix in this system containing simultaneously both a potential well and a barrier waveguide? 
%In the presence of both a barrier and a well, the states with negative and positive dispersion can over lap.
%The gappless nature of graphene's spectrum results in the coexistence of barrier and well 
%both postive and negative dipsersoin, if you want to control the behaavior we need to create both type of carriers, it makes sensse to consider bipolar waveguides, and recombination of particles, correspondong.

Transmission through single and multiple barriers in graphene has been a subject of extensive research~\cite{katsnelson2006chiral,cheianov2006selective,williams2007quantum,cheianov2007focusing,williams2007quantum,pereira2007graphene,huard2007transport, ozyilmaz2007electronic,pereira2008resonant, gorbachev2008conductance,pereira2010klein,bahlouli2012tunneling,alhaidari2012relativistic,wei2013resonant} including periodic potentials~\cite{brey2009emerging} and sinusoidal multiple-quantum-well systems~\cite{xu2010resonant,pham2015tunneling}. 
%Prorogation along asymmetric waveguides~\cite{he2014guided, xu2016guided} formed by different depth square-well have also been considered.
Despite this significant body of research, the phenomenon of pseudo-gap formation in bipolar waveguides has been hitherto overlooked. 
%arising from electron-hole repulsion has been completely overlooked.  
It should be emphasized that the idea of using bipolar waveguides stems directly from the essential feature of graphene, as a gapless material, that a potential barrier can contain guided electron modes, effectively acting as a potential well. For non-relativistic particles, the probability of tunneling between two wells results in the splitting of energy levels, to form a doublet state. In contrast, the probability of tunneling between a well and a barrier in graphene results in the bands forming an avoided crossing at finite $k_y$, where $\hbar k_y$, is the momentum along the waveguide. As we demonstrate below, by modulating the applied voltage, bipolar waveguides have fully controllable pseudo-gaps, which exhibit extremely strong optical transitions. Not only this, but these transitions occur in the highly elusive and desirable THz frequency range. This part of the electromagnetic spectrum is notoriously difficult to generate and manipulate~\cite{lee2007searching}. Therefore, by using a suitably chosen combination of guiding potentials, one can transform a graphene sheet into a narrow-gap nanotube without rolling, where the effective nanotube radius is controlled by the strength of the applied potential.

\begin{figure}[ht]
    \centering
    %size 85 a)
    \includegraphics[width=0.4\linewidth]{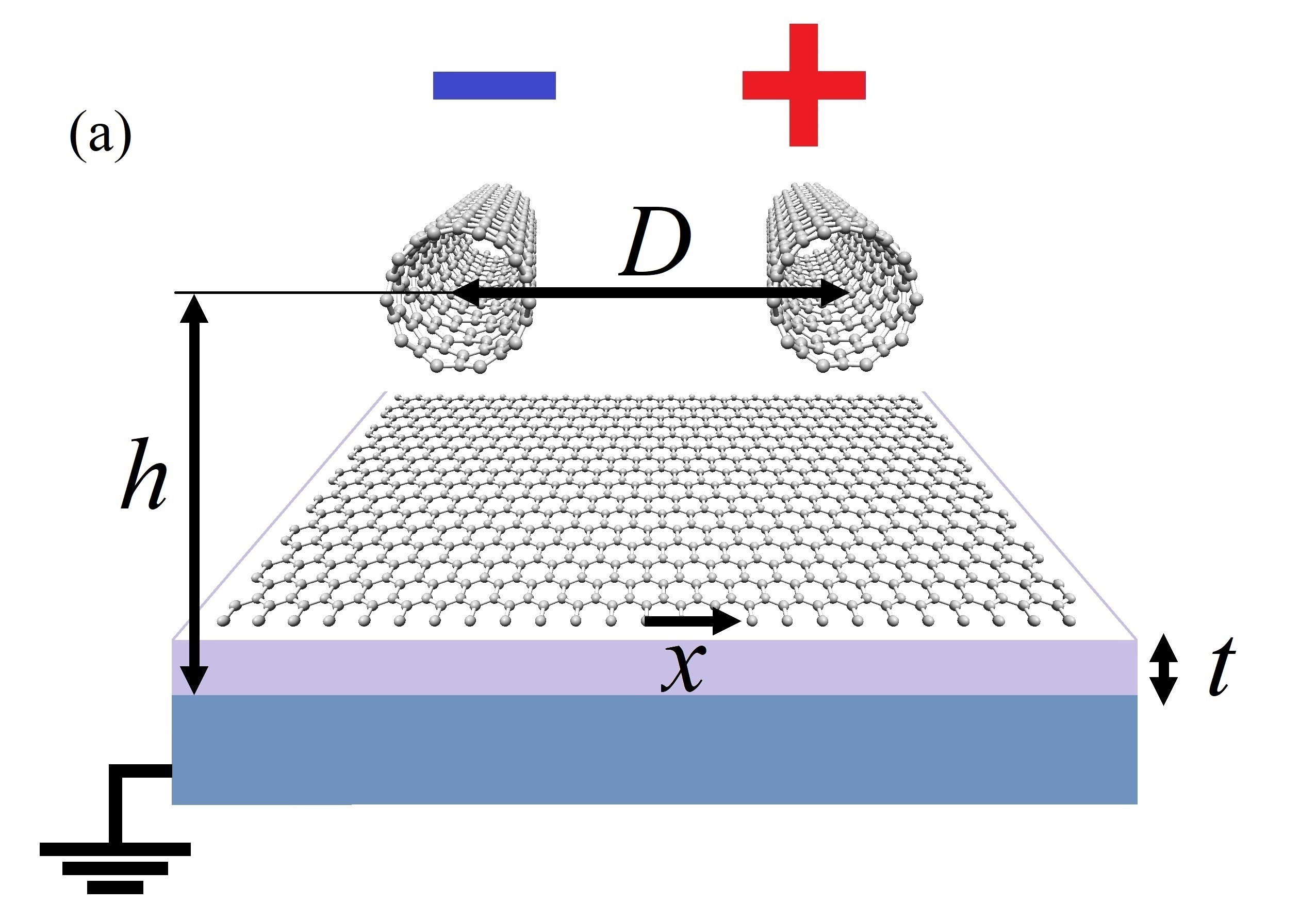}
    \includegraphics[width=0.4\linewidth]{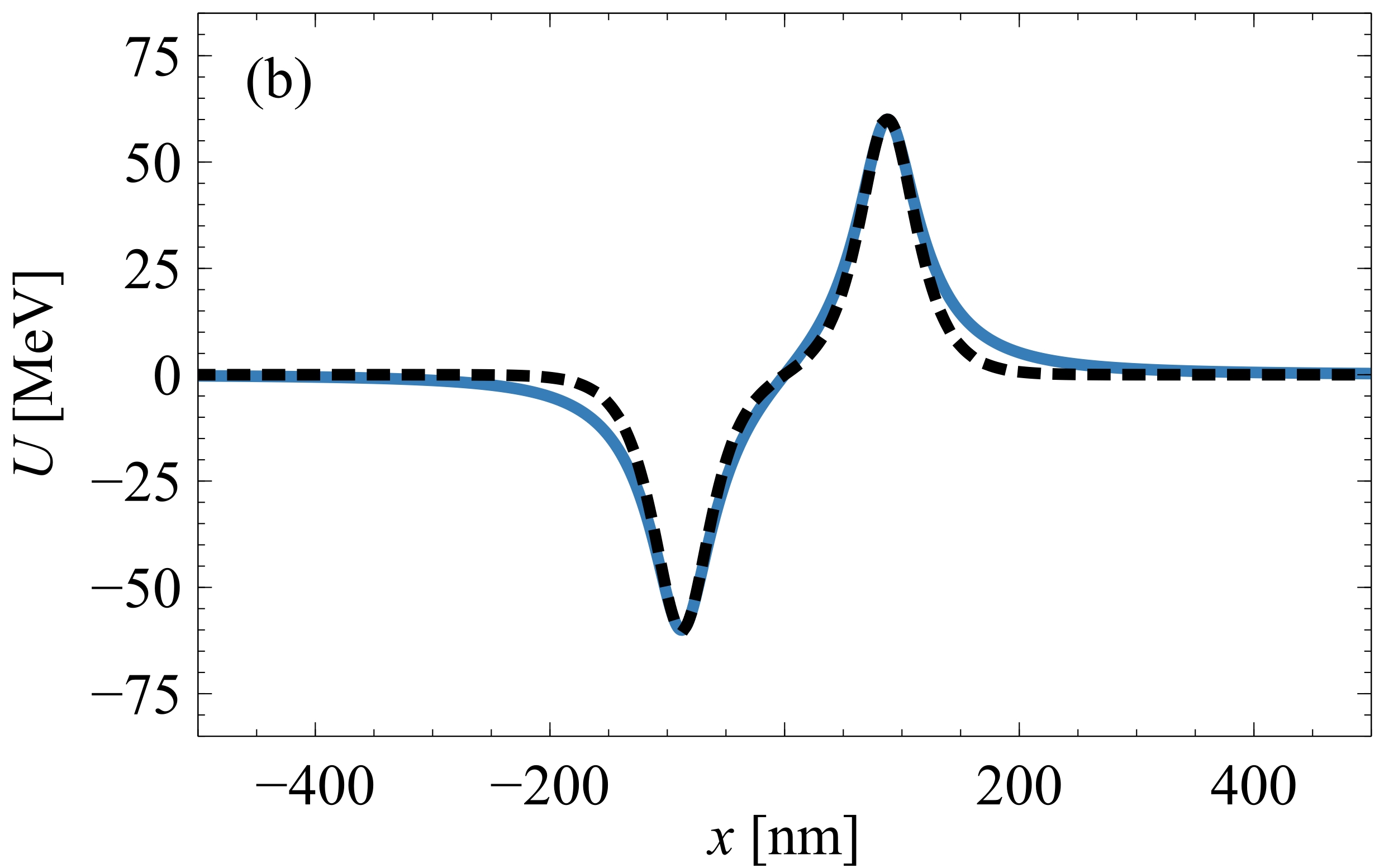}
    \caption{
    (a) The schematic of the proposed experimental setup and (b) a comparison between the potential created by two nanotubes, separated a distance $D=175$~nm, %with the nanotubes and graphene sheet suspended a height, 
    with $h=40$~nm, and $t=20$~nm %above the metallic substrate respectively 
    (grey line) and the linear combination of a well and barrier defined by shifted $\pm u_0 /\cosh(x/L)$ functions (black dashed line) for the case of $\phi_1=-\phi_2=0.25$~V, matching both their peak values and second-derivative at their maxima.}
    \label{fig:set_up}
\end{figure}
Inspired by the most advanced experimentally attainable waveguides~\cite{cheng2019guiding}, our proposed bipolar waveguide is defined by two nanotubes (top gates), of radius $r_{0}$, separated by a distance $D$, as shown in Fig.~\ref{fig:set_up}~(a). Both nanotubes are placed at a height, $h$, above the metallic substrate. The potential profile in the graphene plane separated from the same substrate by another distance $t$ is given by the expression:
\begin{equation}
U\left(x\right)=\frac{e\widetilde{\phi}_{1}}{2}\ln\left[\frac{\left(x+\frac{D}{2}\right)^{2}+\left(h-t\right)^{2}}{\left(x+\frac{D}{2}\right)^{2}+\left(h+t\right)^{2}}\right]+\frac{e\widetilde{\phi}_{2}}{2}\ln\left[\frac{\left(x-\frac{D}{2}\right)^{2}+\left(h-t\right)^{2}}{\left(x-\frac{D}{2}\right)^{2}+\left(h+t\right)^{2}}\right],
\end{equation}
where $\widetilde{\phi}_{1,2}=\phi_{1,2}/\ln\left(\frac{2h-r_{0}}{r_{0}}\right)$, and $\phi_{1}$ ($\phi_{2}$) is the applied voltage between the left (right) nanotube and the back gate. The expression above can be easily modified when the top gates are fully embedded in a dielectric material. In the above-mentioned recent work~\cite{cheng2019guiding}, a smooth electron waveguide was fabricated using a carbon nanotube as a top gate and the graphene sheet was sandwiched in between two layers of hexagonal boron nitride (h-BN). The top h-BN layer had a thickness, $h$, of between 4 and 100 nm, and the bottom layer had a thickness, $t$ of around 20 nm. It can be seen from Fig.~\ref{fig:set_up}~(b) that the potential in the plane of the graphene sheet, derived using image charges in the substrate, can be very well approximated by a linear combination of two shifted hyperbolic secant functions. It is convenient to use this particular approximation because for a single well, the hyperbolic secant potential possesses quasi-exact solutions to the Dirac equation~\cite{hartmann2010smooth,hartmann2011excitons,hartmann2014quasi}. This will enable us to treat both the size of the pseudo-gap and the optical transitions across it analytically.
%Since dielectric constant of h-BN ranges from 3.29 to 3.76\cite{laturia2018dielectric}, and a typical tube diameter if of the order of 1 to 3 nm, 

\section{Negative dispersion and pseudo-gaps}
%\section{Energy level repulsion}
%The best understanding of the physics of the system can be achieved for well separated channels and in this...wells separated equal to the... Let us introduce a 
The low-energy quasiparticle behaviour in graphene is known to be described with spectacular accuracy by the 2D Dirac equation for massless fermions~\cite{wallace1947band}. In the presence of a confining electrostatic potential, $U(x)$, the effective 1D matrix Hamiltonian for confined modes in a graphene waveguide can be written in the standard basis of graphene's two sub-lattices as
\begin{equation}
\hat{H}=\hbar v_{\mathrm{F}}
\left(
\hat{k}_{x}\sigma_{x}+s_{\mathrm{K}}k_{y}\sigma_{y}\right) +\mathrm{I}U(x),
\label{eq:Ham_Dirac}
\end{equation}
where $\hat{k}_{x}=-i\frac{\partial}{\partial x}$, $k_{y}$ is a wavenumber corresponding to the motion along the waveguide, $\sigma_{x,\,y,\,z}$ are the Pauli spin matrices, $\mathrm{I}$ is the 2 by 2 unit matrix, $v_{\mathrm{F}}$ is the Fermi velocity, which is approximately $\approx10^{6}$ m/s, and $s_{\mathrm{K}}$ is the valley quantum number, which has the value of $+1$ and $-1$ for the K and K' valley respectively. 

We are interested in the situation when $U(x)$ is a combination of two fast decaying potentials separated by a distance $d$. For a better understanding of the underlying physics it is instructive to look both at the same and different sign constituent potentials. This allows a comparison with the familiar non-relativistic results for the double quantum well. The results are especially transparent for a quasi-one dimensional potential formed by a combination of either; a well and a barrier of the same strength, or, two equal wells:
\begin{equation}
U\left(x\right)=
u\left(x+\frac{d}{2}\right) \pm  
u\left(x-\frac{d}{2}\right),
\label{eq:pot_to_solve}
\end{equation}
where, $u\left(x\right)$ is an individual symmetric potential well, for which the Hamiltonian, Eq.~(\ref{eq:Ham_Dirac}), admits exact zero-energy eigenfunctions, which we denote as $\Psi_{0}\left(x\right)$. The function $\Psi_{0}\left(x\right)$ is normalized and assumed to rapidly decay outside of the well. The tunneling between wells, or between a well and a barrier, results in the energy level, $E=0$, splitting into two levels, $E_{1}$ and $E_{2}$. At $E=0$, the barrier wave function, denoted $\Psi_{-}$, is the complex conjugate of the well wave function, $\Psi_{-}=\Psi_{0}^{\star}\left(x-\frac{d}{2}\right)$, reflecting the fact that the barrier for electrons is a well for holes; whereas, for the two-well case $\Psi_{+}=\Psi_{0}\left(x-\frac{d}{2}\right)$.
%\textbf{In the small tunneling probability regime}
%where the $\pm$ subscripts corresponds to the sign in equation 3.
In the weak wavefunction overlap approximation, resembling the tight-binding (H\"{u}ckel molecular orbital) methods widely used in solid-state and molecular physics, we may write the wave functions corresponding to eigenvalues $E_1$ and $E_2$ as:
\begin{equation}
\Psi_{1}=\frac{1}{\sqrt{2}}\left[\Psi_{\pm}\left(x-\frac{d}{2}\right)+\Psi_{0}\left(x+\frac{d}{2}\right)\right],
\label{eq:Psi_E_1}
\end{equation}
\begin{equation}
\Psi_{2}=\frac{1}{\sqrt{2}}\left[\Psi_{\pm}\left(x-\frac{d}{2}\right)-\Psi_{0}\left(x+\frac{d}{2}\right)\right].
\label{eq:Psi_E_2}
\end{equation}
Following an approach similar to the non-relativistic case~\cite{landau2013quantum}, but for a matrix Hamiltonian, the energy level splitting, $E_g=\left|E_2-E_1\right|$, can be shown to be
%Using Eq.~(\ref{eq:Ham_Dirac}), Eq.~(\ref{eq:Psi_E_1}) and Eq.~(\ref{eq:Psi_E_2}) the energy level splitting can be shown to be
\begin{equation}
%E_{2}-E_{1}=-2i\Psi_{\pm}^{\dagger}\left(-\frac{d}{2}\right)\sigma_{x}\Psi_{0}\left(\frac{d}{2}\right)
%E_g=2\hbar v_{\mathrm{F}} \left|\Psi_{\pm}^{\dagger}\left(-\frac{d}{2}\right)\sigma_{x}\Psi_{0}\left(\frac{d}{2}\right)\right|.
E_{g}=2\hbar v_{\mathrm{F}}\left|\Psi_{\pm}^{\dagger}\left(-\frac{d}{2}\right)\sigma_{x}\Psi_{0}\left(\frac{d}{2}\right)\right|.
\label{eq:gap_guess}
\end{equation}
It should be noted that there is a striking difference between the relativistic and non-relativistic case. For the non-relativistic case the splitting is proportional to the product of the single well function and its derivative~\cite{landau2013quantum}; whereas, in the relativistic case, it depends only on the individual well and barrier functions (or the two shifted well functions for the double well). For simplicity, we considered above two potentials of equal strengths, and estimated the splitting of the $E=0$ state. These results can be easily generalized for non-equal potentials and non-zero values of energies as long as the energy levels in the individual potentials coincide~(see Appendix). This theorem demonstrates the utility of quasi-exact solutions to the Dirac equation~\cite{hartmann2010smooth,hartmann2014quasi,hartmann2017two} for bipolar waveguides. Indeed, the exact solutions often correspond to the case where there is symmetry between the positive and negative energy solutions, allowing all pseudo-gaps to be treated within this formalism. Furthermore, knowledge of the exact wave-functions allows the matrix element of optical transitions across the pseudo-gaps to be calculated analytically.

\begin{figure}[ht]
    \centering
    \includegraphics[width=0.5\linewidth]{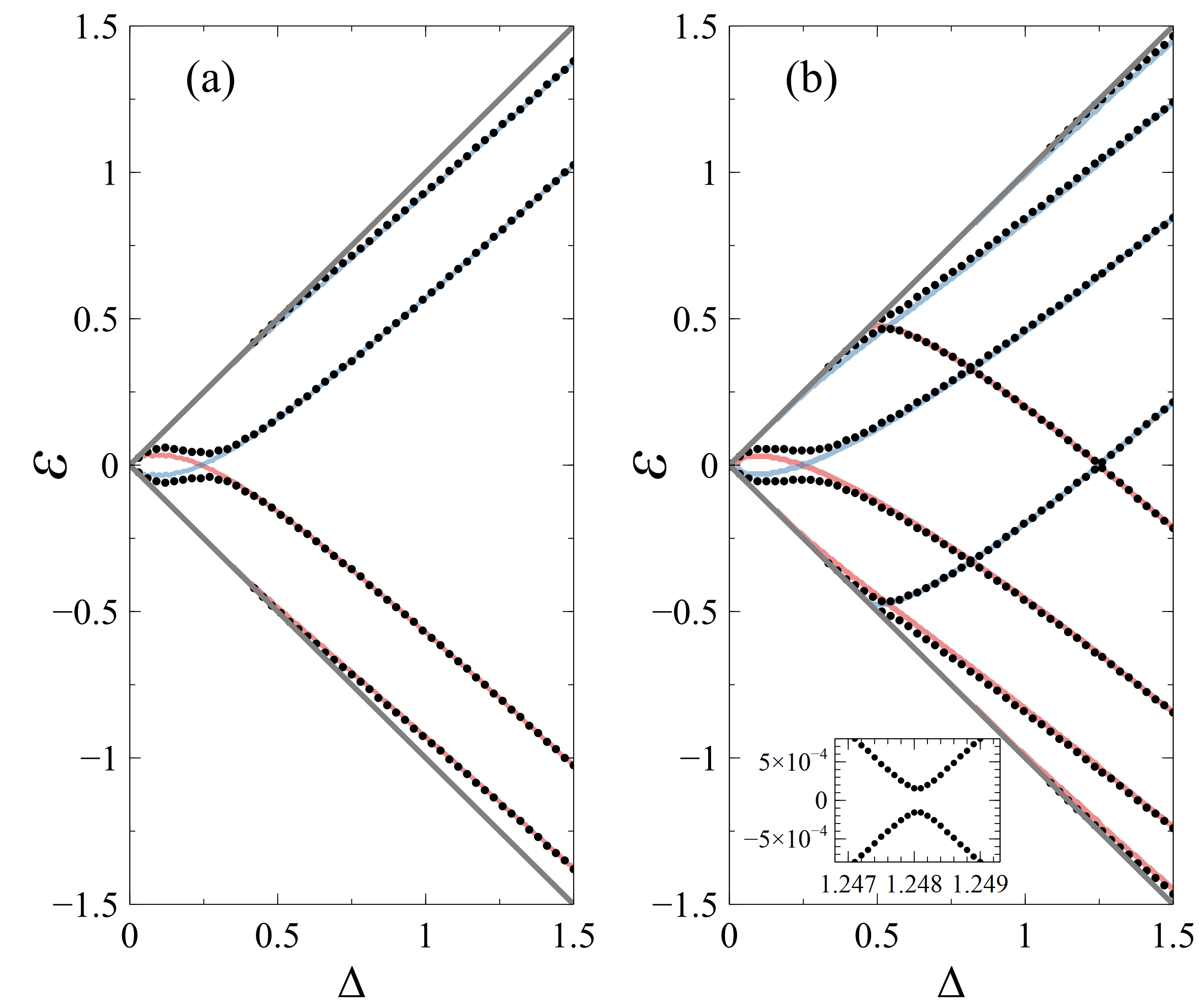}
    \includegraphics[width=0.5\linewidth]{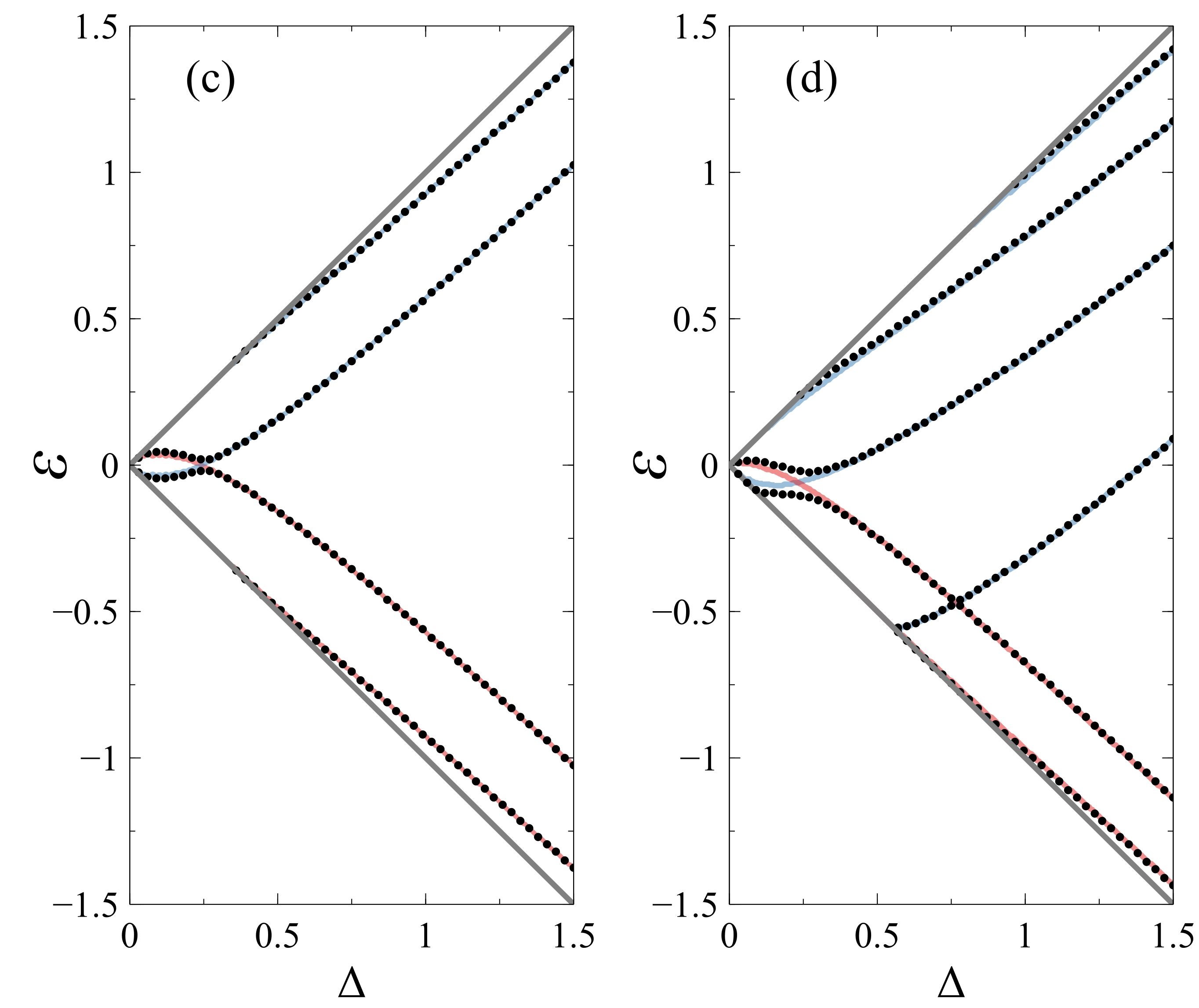}
    \caption{
The energy spectrum in dimensionless units, $\varepsilon=EL/(\hbar v_{\mathrm{F}})$ \textit{vs} $\Delta=\left|k_y\right|L$, of confined states in a bipolar waveguide (shown as black dotted lines), defined by the potential $-u_{1}/\cosh\left[\left(x+\frac{d}{2}\right)/L\right]+u_{2}/\cosh\left[\left(x-\frac{d}{2}\right)/L\right]$, for (a) $u_1 L =u_2 L =0.75\,\hbar v_{\mathrm{F}}$, $d/L=8$, (b) $u_1 L=u_2 L =1.75\,\hbar v_{\mathrm{F}}$, $d/L=8$, (c) $u_{1} L =u_{2} L =0.75\,\hbar v_{\mathrm{F}}$, $d/L=12$ and (d) $u_{1} L =1.9\, \hbar v_{\mathrm{F}}$, $u_{2} L =0.6\, \hbar v_{\mathrm{F}}$, $d/L=8$. The blue and red lines show the dispersion lines for an isolated well and barrier respectively, while the grey lines show the boundary at which the bound states merge with the continuum at $\left|E\right|=\hbar v_{\mathrm{F}} \left|k_y\right|$.
%The energy spectrum in dimensionless units, $\varepsilon=E/(\hbar v_{\mathrm{F}}/L)$ \textit{vs} $\Delta=\left|k_y\right|L$, of confined states in a bipolar waveguide (shown as black dotted lines), for a tube separation of $d/L=8$ and a potential strength of (a) $\omega=u_0/(\hbar v_{\mathrm{F}}/L)=0.75$ and (b) $\omega=1.75$. The blue and red lines show the dispersion lines for an isolated well and barrier respectively. The inset shows the detailed view of the second (for a larger value of $\Delta$) zero-energy gap opening. The boundary at which the bound states merge with the continuum, $\left|E\right|=\hbar v_{\mathrm{F}} \left|k_y\right|$, are denoted by the grey lines.
%The energy spectrum in dimensionless units, $\varepsilon=E/(\hbar v_{\mathrm{F}}/L)$ \textit{vs} $\Delta=\left|k_y\right|L$, of confined states in a bipolar waveguide (shown as black dotted lines), defined by the potential $-u_{1}/\cosh\left[\left(x+\frac{d}{2}\right)/L\right]+u_{2}/\cosh\left[\left(x-\frac{d}{2}\right)/L\right]$, for the case of (a) $u_{1}=u_{2} =0.75\,\hbar v_{\mathrm{F}}/L$, $d/L=12$ and (b) $u_{1} =1.9\, \hbar v_{\mathrm{F}}/L$, $u_{2} =0.6\, \hbar v_{\mathrm{F}}/L$, $d/L=8$. The blue and red lines show the dispersion lines for an isolated well and barrier respectively, while the grey lines show the boundary at which the bound states merge with the continuum at $\left|E\right|=\hbar v_{\mathrm{F}} \left|k_y\right|$.
}
%step := 0.1/5000;step_M := 0.01/100;M_Val_Max := 1.252;M_Val_Min:= 1.244;
    \label{fig:energy_fixed_dist}
\end{figure}
%The Klein tunneling effect means that when choosing the appropriate well and barrier model potential, one should select only those which decay to zero at infinity. Else, outside the well/barrier there will be an infinite number of states to tunnel into. Besides this, physical top-gates generate potentials vanish at infinity, and potentials which vary slowly over the graphene lattice constant do not result in inter-valley mixing. 
%To maximize the full strength of the analytic expression above
%In additional to providing an excellent fit to a realistic potential

In what follows we shall model a bipolar waveguide as
%a linear combination of the potential
\begin{equation}
U(x)=-\frac{u_{1}}{\cosh\left[\left(x+\frac{d}{2}\right)/L\right]}+\frac{u_{2}}{\cosh\left[\left(x-\frac{d}{2}\right)/L\right]},
\label{eq:potential}
\end{equation}
where $u_1$ and $u_2$ are the depth of the well and the height of the barrier, respectively, and $L$ is the effective width of the potential, which for now we assume to be the same for the well and the barrier. For $u_1=u_2=u_0>0$, this smooth potential indeed provides an excellent fit to the realistic potential generated by two oppositely charged nanotubes above the surface of graphene, see Fig.~\ref{fig:set_up}~(b). Also each of the individual secant potentials supports exact analytic solutions at $E=0$~\cite{hartmann2010smooth}, allowing the comparison of the numerical results with the approximate formula, Eq.~(\ref{eq:gap_guess}). It is convenient to introduce similar dimensionless parameters as Ref.~\cite{hartmann2010smooth}, namely $\omega=u_{0} L / ( \hbar v_{\mathrm{F}})$ and $\Delta=\left|k_{y}\right| L$. It should be noted that the number of bound states contained within a realistic confining potential is also defined by the product of the characteristic potential depth and its width \cite{cheng2019guiding,hartmann2017two} rather than its exact form. Effects such as non-linear screening and the renormalization of the Fermi velocity~\cite{elias2011dirac}, can be accommodated within the same dimensionless parameter, $u_0 L/(\hbar v_F)$. The particular choice of the potential we use does not influence the physical picture. The use of this dimensionless parameter also allows the application of our results to other 2D systems with linear dispersion beyond graphene, e.g., surface states of topological insulators.

In the absence of inter-potential tunneling, the dispersion lines of the well and barrier (indicated in Fig.~\ref{fig:energy_fixed_dist} by blue and red lines respectively) cross when $\Delta_n=\omega - n -1/2$, where $n$ is a positive integer~\cite{hartmann2010smooth}. The corresponding exact wavefunctions when substituted into Eq.~(\ref{eq:gap_guess}) yield the following approximate expression for the $n=0$ pseudo-gap (see Appendix)
%\begin{equation}
%   E_{g}=\frac{2\hbar %v_{\mathrm{F}}e^{-\left(\omega-\frac{1}{2}\right)\frac{d}{L}}}{LB\left(\omega+1/2,\omega-1/2\right)},
%\end{equation}
\begin{equation}
E_{g}/\left(\hbar v_{\mathrm{F}}/L\right)\approx\frac{2\exp\left(-\Delta_{0}d/L\right)}{B\left(1+\Delta_{0},\Delta_{0}\right)},
%E_{g}=\frac{2\hbar v_{\mathrm{F}}e^{-\left(\omega-\frac{1}{2}\right)\frac{d}{L}}}{LB\left(\omega+1/2,\omega-1/2\right)},
%E_{g}\approx\frac{2\hbar v_{\mathrm{F}}e^{-\alpha_{0}\Delta_0}}{LB\left(1+\Delta_0,\Delta_0\right)},
\label{eq:approx_ana_gap}
\end{equation}
where $B(m,n)$ is the Beta function. This formula gives an extremely good approximation to the numerical solution in the limit when the ratio between the wire separation and the effective width of the potential is large, i.e., $d/L \gg 1$. 

%, which must exceed $? \pi$ for the small tunneling approximation to be valid.
%should be emphasised that this approx works for well separated wells, $d\gg L$ and deep enough wells, Delta>>0

In Fig.~\ref{fig:energy_fixed_dist} we plot the numerically obtained energy dependence on $\Delta$, i.e. the momentum along the barrier in dimensionless units, for the cases of $\omega=0.75$ (panel~(a)) and $\omega=1.75$ (panel~(b)). In both cases the two oppositely charged nanotubes are separated by a distance $d/L=8$, which corresponds to approximately $175$~nm for the case of $h=40$~nm and $t=20 $~nm. 
At this distance the energy-level splitting formula, Eq.~(\ref{eq:gap_guess}), accurately predicts the value of the $n=0$ pseudo-gap within a few per cent error. This error becomes one order of magnitude smaller when $d/L=12$. It is instructive to compare these electrostatically induced pseudo-gaps with curvature-induced gaps in carbon nanotubes. For a narrow-gap carbon nanotube, it is well established that the larger is its radius, the smaller is the curvature-induced gap~\cite{kane1997size}. Therefore, the strength of the applied voltage, for a particular guided mode, can be mapped to the radius of the nanotube. Increasing the voltage results in more tightly confined guided modes, characterised by a higher value of $\Delta_0$ entering Eq.~(\ref{eq:approx_ana_gap}), therefore we arrive to a smaller value of the pseudo-gap, which moves to the right in Fig.~\ref{fig:energy_fixed_dist}~(b).
%thus decreasing the overlap between the two wavefunctions, resulting in a smaller value of the pseudo-gap. 
%Concurrently, the avoided crossing will shift to increasing values of $\Delta$ as the applied voltage increases, since zero-energy bound states are subject to the condition $\Delta_n=\omega - n -1/2$. 
It can also be seen from panel (b) of this figure that the deeper the well and higher the barrier, the more states are contained within each channel, increasing correspondingly the number of avoided crossings appearing in the dispersion. It should also be noted that although increasing the voltage leads to stronger confinement for lower order modes, it also results in the appearance of higher order modes which are more spread out, leading to additional wider pseudo-gaps. In Fig.~\ref{fig:energy_fixed_dist}~(b), which corresponds to the case of a deeper well and higher barrier, we can see two pseudo-gaps at $E=0$ as well as additional pseudo-gaps at non-zero energy since there are more guided modes in the low-energy part of the spectrum.
%enables an increase in the over lap between the well and barrier functions. 

%n=0 mode
%Fig 2  a) n=0, predicted 72.937880818487850075 meV; 76 mev

%       b) n=0, predicted 0.2933957206 meV;  0.314 mev

%       b) n=1, predicted 92.27089314  meV;  92 meV

%Fig 3 a) n=0, predicted 26.854446142376919353 meV; 26.736 mev

%\begin{figure}[h]
%    \centering
%    \includegraphics[width=0.7\linewidth]{Figure_3.jpg}
%    \caption{
%}
%    \label{fig:energy_fixed_volt}
%\end{figure}

It can be seen by comparing Fig.~\ref{fig:energy_fixed_dist}~(c) to~(a) that increasing the distance between the nanotubes, decreases the size of the gap. This is a result of the decrease in the overlap between the well and barrier functions. Technologically it is quite difficult to have exact control over the precise tube separation. However, this is not so important since it is possible to control the value of the pseudo-gap by the applied voltage. Furthermore, it can be see from Fig.~\ref{fig:energy_fixed_dist}~(d) that the effect of pseudo-gap opening is robust against asymmetry in the system. In Fig.~\ref{fig:energy_fixed_dist}~(d), the dispersion is recalculated for two tubes separated at the same distance as in Fig.~\ref{fig:energy_fixed_dist}~(a), but with the depth of the well increased to $\omega=1.9$, while the size of the barrier decreased to $\omega=0.6$. This demonstrates that top gates of mismatched radius, or dissimilar magnitudes of applied voltage will just shift the value in energy in which the avoided crossing occurs. Although Eqs.~(\ref{eq:gap_guess},\ref{eq:approx_ana_gap}) give for the double well and bipolar waveguide the same value of energy-level splitting at the value of $k_y$ corresponding to $E=0$ in a single well, for the double well there is no pseudo-gap and the dispersion remains monotonous, this case is considered in depth elsewhere~\cite{Hartmann2020double}.

\section{Interband Transitions}
In what follows we shall demonstrate that much like in narrow-gap nanotubes~\cite{hartmann2019interband}, the %reconstruction of the 
wavefunction intermixing leads to strongly allowed optical transitions across the pseudo-gaps. %associated with these avoided crossings. 
%In the presence of an electromagnetic field, the particle momentum operator, $\hat{\boldsymbol{p}}$, is modified such that $\hat{\boldsymbol{p}}\rightarrow\hat{\boldsymbol{p}}+e\boldsymbol{A}/c$. 
The probability of optical transitions is proportional to the squared modulus of the matrix element of velocity operator between the relevant states. The velocity operator written in the same basis as the Hamiltonian given in Eq.~(\ref{eq:Ham_Dirac}) is~\cite{saroka2018momentum,hartmann2019interband} 
%The general form of the perturbation due to an electromagnetic wave impinging normally to a Dirac material is
\begin{equation}
\hat{\boldsymbol{v}}=v_{\mathrm{F}}\left(\sigma_{x}\hat{\boldsymbol{x}}+s_{\mathrm{K}}\sigma_{y}\hat{\boldsymbol{y}}\right).
\label{eq:VME}
\end{equation}
The probability of a dipole transition is proportional to $\left|\left\langle \Psi_{f}\left|\hat{\boldsymbol{v}}\cdot\textbf{e}\right|\Psi_{i}\right\rangle \right|^{2}$,
where $\Psi_{i}$ and $\Psi_{f}$ are the initial and final states, respectively, and $\textbf{e}=(e_x,e_y)$ is the light polarization vector. For linearly polarized light, $\textbf{e}=(\cos\left(\varphi_{0}\right),\sin\left(\varphi_{0}\right))$, while for right- and left-handed polarized light $\textbf{e}=\left(1,-i\right)/\sqrt{2}$ and $\textbf{e}=\left(1,i\right)/\sqrt{2}$. Within the small wavefunction overlap approximation, for a bipolar waveguide defined by the potential given by Eq.~(\ref{eq:potential}) with $u_1=u_2=u_0>0$, the matrix element of velocity of the $n=0$ mode at $\Delta=\Delta_0$ is~(see Appendix)
\begin{equation}
\left|\left\langle \Psi_{2}\left|\hat{\boldsymbol{v}}\right|\Psi_{1}\right\rangle \right|/v_{\mathrm{F}}\approx\left|\frac{e_{x}B\left(1-e^{-d/L};\frac{1}{2}+\Delta_{0},\,0\right)e^{-\Delta_{0}d/L}}{B\left(1+\Delta_{0},\,\Delta_{0}\right)}-i\frac{k_{y}}{\left|k_{y}\right|}\frac{e_{y}B\left(\frac{1}{2}+\Delta_{0},\,\frac{1}{2}+\Delta_{0}\right)}{B\left(1+\Delta_{0},\,\Delta_{0}\right)}\right|,
%\left|\left\langle \Psi_{2}\left|\hat{\boldsymbol{v}}\right|\Psi_{1}\right\rangle /v_{\mathrm{F}}\right|\approx\left|\frac{e_{x}B\left(1-e^{-d/L};\frac{1}{2}+\Delta_{0},\,0\right)e^{-\Delta_{0}d/L}}{B\left(1+\Delta_{0},\,\Delta_{0}\right)}-is_{\mathrm{K}}\frac{e_{y}B\left(\frac{1}{2}+\Delta_{0},\,\frac{1}{2}+\Delta_{0}\right)}{B\left(1+\Delta_{0},\,\Delta_{0}\right)}\right|,
%\left\langle \Psi_{2}\left|\hat{\boldsymbol{v}}\right|\Psi_{1}\right\rangle /v_{\mathrm{F}}\approx-s_{\mathrm{K}}\frac{e_{y}B\left(\frac{1}{2}+\Delta_{0},\,\frac{1}{2}+\Delta_{0}\right)}{B\left(1+\Delta_{0},\,\Delta_{0}\right)}-i\frac{e_{x}B\left(1-e^{-d/L};\frac{1}{2}+\Delta_{0},\,0\right)e^{-\Delta_{0}d/L}}{B\left(1+\Delta_{0},\,\Delta_{0}\right)},
\label{eq:n_0_vme}
\end{equation}
where $B(x;m,n)$ is the incomplete Beta function. For the case of the double well, at the same value of momentum, the matrix element of velocity of the $n=0$ mode is the first term of Eq.~(\ref{eq:n_0_vme}), whereas $v_y=0$. It reflects the fact that transitions in a double well are only caused by light polarized along the $x$-direction (normal to the waveguide), similar to the non-relativistic case and the 1D square-well potential in graphene~\cite{avishai2020klein}.

\begin{figure}[th]
    \centering
    \includegraphics[width=0.55\linewidth]{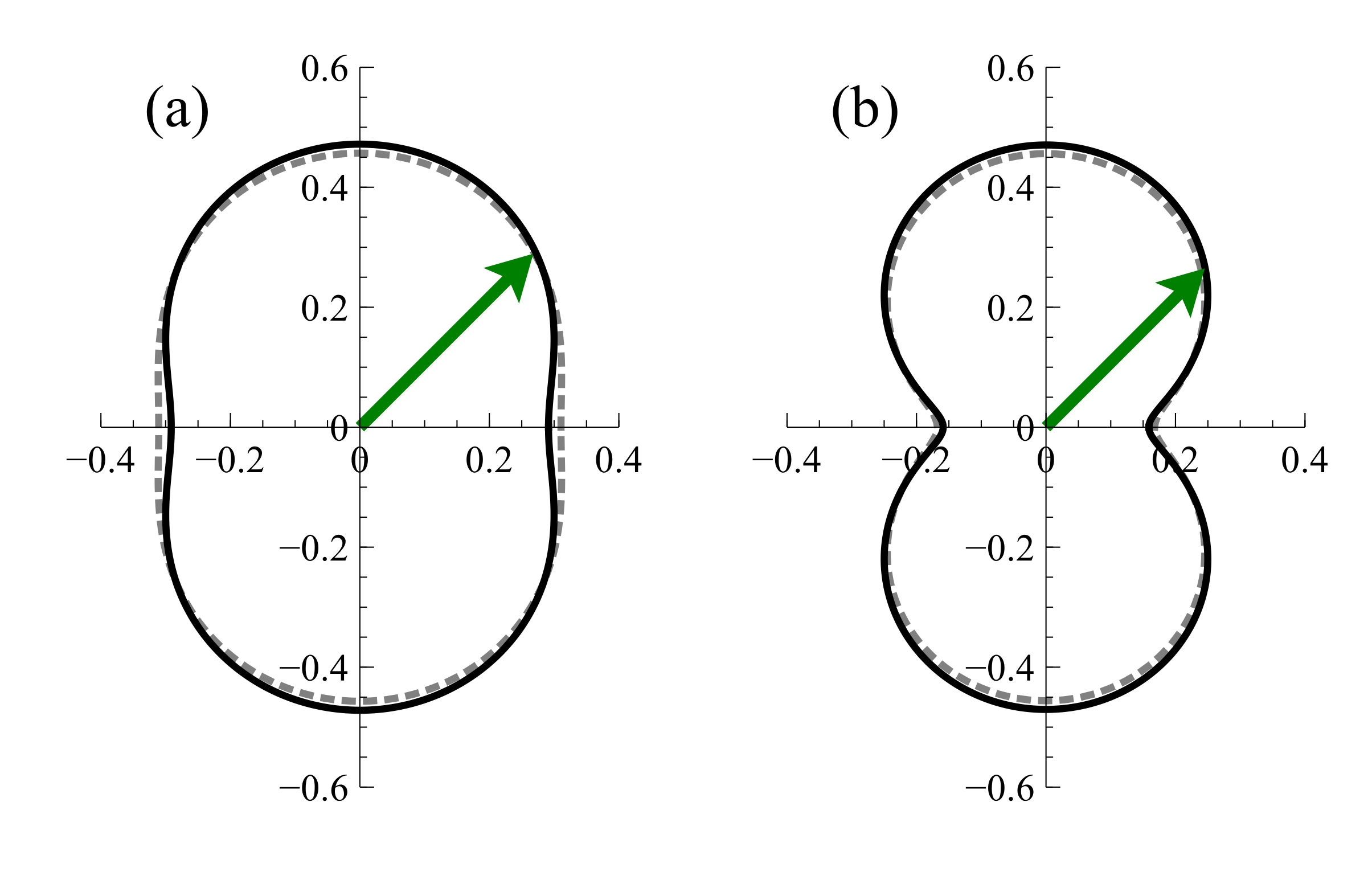}
    \caption{
Polar plots, showing the dependence of the absolute value (the length of the green arrow) of the velocity matrix element (in units of $v_{\mathrm{F}}$) across the $n=0$ pseudo-gaps for dipole transitions caused by normally-incident linearly polarized light on the angle between the polarization vector and the $x$-axis normal to the bipolar waveguide, defined by $\omega=0.75$, for two different top gate separations: (a) $d/L=8$ and (b) $d/L=12$. The analytic approximation is depicted by the dashed grey line and the numerically obtained values are shown by the black solid line.
%Polar plots, whose magnitude (depicted for example by the length of the green arrows) represents the absolute value of the matrix element of velocity (in units of $v_{\mathrm{F}}$), $\left|\left\langle \Psi_{2}\left|\hat{\boldsymbol{v}}\right|\Psi_{1}\right\rangle /v_{\mathrm{F}}\right|$, as a function of linear polarization angle (depicted by the arrow's angle relative to the $x$-axis), for transitions across the $n=0$ pseudo-gap edge for a bipolar waveguide defined by $\omega=0.75$ for two different top gate separations; a) $d/L=8$ and b) $d/L=12$. These pseudo-gaps can be seen in Fig.~\ref{fig:energy_fixed_dist}~(a) and Fig.~\ref{fig:energy_fixed_volt}~(a) respectively. The analytic approximation is depicted by the dashed grey line and the numerically obtained values are shown by the black solid line.}
\label{fig:vme}
}
\end{figure}

In stark contrast, the transition across the pseudo-gap of a bipolar waveguide is strongly polarized along the $y$-axis (waveguide direction). The situation changes away from the pseudo-gap. For small values of $\left|k_{y}\right|$, the transitions are polarized normally to the $y$-direction, as expected from the momentum alignment phenomenon in graphene~\cite{saroka2018momentum}. For large values of $\left|k_{y}\right|$, the overlap between the well and barrier wavefunctions becomes very small leading to vanishing transition probabilities for both polarizations. This effect very much resembles the situation in a narrow-gap carbon nanotube, where optical transitions polarized along the nanotube axis are allowed in the narrow energy interval around the curvature-induced gap~\cite{hartmann2019interband}. The main difference from a nanotube can be clearly seen from Fig.~\ref{fig:vme}, namely both transitions polarizations along and normal to the waveguide are allowed across the pseudo-gap. The contribution of the $x$-component exponentially decreases with an increase in top gate potential or the separation between the gates, as can be seen from Fig.~\ref{fig:vme}. 

The presence of both polarizations for pseudo-gap transitions leads to an effect absent in both graphene and non-chiral nanotubes. Namely, right- and left-handed polarized light produces different populations of pseudo-valleys with opposite signs of $k_y$. This can be clearly seen from Eq.~(\ref{eq:n_0_vme}).  This feature does not depend on the model describing the potential~(see Appendix).

The discussed transitions across the pseudo-gap, fully controlled by the top gate voltages, can be easily brought into the highly desirable THz frequency range, which is usually extremely difficult to control. The presence of the van Hove singularity at the pseudo-gap edge enhances the strength of these transitions. These effects opens the avenue for novel gate-controlled polarized sensitive THz detectors based on bipolar waveguides in graphene.

%\textbf{This offers an additional degree of control, for example, when the Fermi level is inside one of the “higher” avoided crossings the bulk THz transition can be blocked by Pauli blocking but the guided mode transition is still allowed and also enhanced by the shifted van Hove singularity.}
%The $n=0$ pseudo-gap in Fig.~\ref{fig:energy_fixed_volt}~(a) is 6.5~THz

\section{Conclusions}
To conclude, we have shown that bipolar waveguides allows for the creation of a non-monotonous 1D dispersion along the electron waveguide. The repulsion of well and barrier states results in the appearance of pseudo-gaps in the spectrum, whose size and symmetry can be fully controlled by the top gate voltages. These gaps can be estimated analytically for exactly-solvable potentials. The opening of these pseudo-gaps results in strongly allowed THz transitions with non-trivial optical selection rules. %for an excitation incident normally to the graphene plane and linearly polarized along the waveguide, as well a valley-dependent optical selection rules for circularly polarized light of the same incidence.
The predicted negative dispersion of the guided modes may lead to various other physical effects ranging from solitary waves~\cite{gulevich2017exploring} to Gunn-diode type current oscillations~\cite{gunn1963microwave}.

%These strongly allowed optical transitions can be tuned to fall within the much sort after THz frequency range. Finally, the highly polarisation sensitive optical selection rules coupled with the presence of the van Hove singularity at the pseudo-gap edge makes graphene bipolar waveguides potential building blocks for polarisation sensitive THz radiation detectors and opto-valleytronic devices~\cite{hartmann2010optoelectronic}.
\section*{Acknowledgements}
We thank C.A. Downing for his critical reading of the manuscript. This work was supported by the EU H2020 RISE project TERASSE (H2020-823878). RRH acknowledges financial support from URCO (71 F U 3TAY18-3TAY19). The work of MEP was supported by the Ministry of Science and Higher Education of Russian Federation, Goszadanie no. 2019-1246. 

\appendix

\section{Derivation of the approximate expression for the pseudo-gap}\label{Append_gap}
In what follows, we show how the bound state eigenfunctions of an individual quasi-1D quantum well in graphene can be used to obtain the approximate size of the pseudo-gaps present in a graphene bipolar waveguide's energy spectrum. Let us consider a bipolar waveguide created by the 1D guiding potential,
\begin{equation}
U\left(x\right)=u_{\mathrm{I}}\left(x+\frac{d}{2}\right)+u_{\mathrm{II}}\left(x-\frac{d}{2}\right),
\label{eq:potential_s}
\end{equation}
built from two separate guiding channels $u_{\mathrm{I}}$ and $u_{\mathrm{II}}$, separated a distance $d$ apart. The single well and barrier 2D Dirac equations are
\begin{equation}
\left[\hat{H}_{0}+\mathrm{I}u_{\mathrm{I}}\left(x+\frac{d}{2}\right)\right]\Psi_{\frac{d}{2}}=E\Psi_{\frac{d}{2}},
\label{eq:Ham_well}
\end{equation}
and
\begin{equation}
\left[\hat{H}_{0}+\mathrm{I}u_{\mathrm{II}}\left(x-\frac{d}{2}\right)\right]\Psi_{-\frac{d}{2}}=E\Psi_{-\frac{d}{2}},
\label{eq:Ham_hump}
\end{equation}
respectively, and $\hat{H}_{0}=\hbar v_{\mathrm{F}}\left(-i\sigma_{x}\frac{\partial}{\partial x}+\sigma_{y}k_{y}\right)$, where $\sigma_{x,y}$ are the Pauli spin matrices, $\mathrm{I}$ is the 2 by 2 unit matrix, and $k_{y}$ is the electron wavenumber along the direction of the waveguide, and $\Psi_{\pm\frac{d}{2}}$ is a two component wavefunction of the form
\begin{equation}
\Psi_{\pm\frac{d}{2}} 
=\frac{1}{N_{\pm\frac{d}{2}}}\left(\begin{array}{c}
\psi_{A}\left(x\pm\frac{d}{2}\right)\\
\psi_{B}\left(x\pm\frac{d}{2}\right)
\end{array}\right),
\label{eq:wave_fun}
\end{equation}
where the subscripts $\frac{d}{2}$ and $-\frac{d}{2}$
correspond to the left and right guiding potential, respectively. Finally, the spinor components $\psi_{A}$ and $\psi_{B}$ are associated with graphene's two sub-lattices and $N^{2}_{\pm\frac{d}{2}}=\int_{-\infty}^{\infty}\left(\left|\psi_{A}\left(x\pm\frac{d}{2}\right)\right|^{2}+\left|\psi_{B}\left(x\pm\frac{d}{2}\right)\right|^{2}\right)dx$. 

%From hereon we consider bipolar waveguides built from symmetric wells and barriers.
%, allowing us to construct the solutions of $\psi_{A}\left(x\pm\frac{d}{2}\right)$ and $\psi_{B}\left(x\pm\frac{d}{2}\right)$ to be purely imaginary and real respectively. 
Let us consider the case when at energy, $E$, the bound state energy level contained within each guiding potential coincides, and splits into the levels $E_{1}$ and $E_{2}$. In this instance their corresponding wavefunctions are
\begin{equation}
\Psi_{1}=\frac{1}{\sqrt{2}}\left(\Psi_{-\frac{d}{2}}+\Psi_{\frac{d}{2}}\right),
\label{eq:Psi_E_1_s}
\nonumber
\end{equation}
\begin{equation}
\Psi_{2}=\frac{1}{\sqrt{2}}\left(\Psi_{-\frac{d}{2}}-\Psi_{\frac{d}{2}}\right),
\label{eq:Psi_E_2_s}
\nonumber
\end{equation}
%$\Psi_{+}$ and $\Psi_{-}$ 
which obey the differential equations
\begin{equation}
\left[\hat{H}_{0}+\mathrm{I}U\left(x\right)\right]\Psi_{1}=E_{1}\Psi_{1},
\label{eq:Ham_plus}
\end{equation}
\begin{equation}
\left[\hat{H}_{0}+\mathrm{I}U\left(x\right)\right]\Psi_{2}=E_{2}\Psi_{2}.
\label{eq:Ham_minus}
\end{equation}
In the small wavefunction overlap approximation, for the rightmost waveguide, the function $\Psi_{\frac{d}{2}}$ is assumed to be vanishingly small in comparison to $\Psi_{-\frac{d}{2}}$ and vice versa, i.e., $\intop_{0}^{\infty}\Psi_{-\frac{d}{2}}^{\dagger}\Psi_{\frac{d}{2}}dx\approx0$. Multiplying the above expressions by $\Psi_{-\frac{d}{2}}^{\dagger}$, subtracting Eq.~(\ref{eq:Ham_plus}) from Eq.~(\ref{eq:Ham_minus}), and integrating from $x=0$ to $\infty$ yields 
\begin{equation}
%-2e^{-i\varphi}\intop_{0}^{\infty}\left[\Psi_{\frac{d}{2}}^{\dagger}\mathrm{I}u_{2}\left(x-\frac{d}{2}\right)\Psi_{-\frac{d}{2}}\right]^{\star}dx=E_{-}-E_{+},
E_{g}=2\left|\intop_{0}^{\infty}\left[\Psi_{\frac{d}{2}}^{\dagger}\mathrm{I}u_{\mathrm{II}}\left(x-\frac{d}{2}\right)\Psi_{-\frac{d}{2}}\right]^{\star}dx\right|,
\label{eq:Ham_hump_2}
\end{equation}
where $E_{g}=\left|E_{2}-E_{1}\right|$. Using Eq.~(\ref{eq:Ham_hump}) and Eq.~(\ref{eq:Ham_hump_2}) allows one to write
\begin{equation}
%2e^{-i\varphi}\intop_{0}^{\infty}\left(\Psi_{\frac{d}{2}}^{\dagger}\hat{H}_{0}\Psi_{-\frac{d}{2}}\right)^{\star}dx=E_{-}-E_{+}.
%\\
E_{g}=2\left|\intop_{0}^{\infty}\left(\Psi_{\frac{d}{2}}^{\dagger}\hat{H}_{0}\Psi_{-\frac{d}{2}}\right)^{\star}dx\right|.
\label{eq:to_int}
\end{equation}
Substituting the wavefunctions, Eq.~(\ref{eq:wave_fun}), into Eq.~(\ref{eq:to_int}), followed by integration by parts yields
%-i2e^{-i\varphi}\hbar v_{\mathrm{F}}\Psi_{-\frac{d}{2}}^{\dagger}\left(0\right)\sigma_{x}\Psi_{\frac{d}{2}}\left(0\right)+2e^{-i\varphi}\hbar v_{\mathrm{F}}\intop_{0}^{\infty}\Psi_{-\frac{d}{2}}^{\dagger}\mathrm{I}u_{1}\left(x+\frac{d}{2}\right)\Psi_{\frac{d}{2}}dx=E_{-}-E_{+}.
\begin{equation}
E_{g}=2\left|\hbar v_{\mathrm{F}}\Psi_{-\frac{d}{2}}^{\dagger}\left(0\right)\sigma_{x}\Psi_{\frac{d}{2}}\left(0\right)+i\intop_{0}^{\infty}\Psi_{-\frac{d}{2}}^{\dagger}\mathrm{I}u_{\mathrm{I}}\left(x+\frac{d}{2}\right)\Psi_{\frac{d}{2}}dx\right|.
\nonumber
\end{equation}
From $x=0$ to $\infty$, the integral appearing in the above expression is negligibly small compared to the lead term. %By constructing the solutions of $\psi_{A}\left(x\pm\frac{d}{2}\right)$ and $\psi_{B}\left(x\pm\frac{d}{2}\right)$ to be purely imaginary and real respectively, enforces that $\phi=0$. 
Therefore, the size of the pseudo-gap, $E_{g}$, can be expressed as
\begin{equation}
E_{g}=2\hbar v_{\mathrm{F}}\left|\Psi_{-\frac{d}{2}}^{\dagger}\left(0\right)\sigma_{x}\Psi_{\frac{d}{2}}\left(0\right)\right|.
\label{eq:energy_gap}
\end{equation}

We shall now show that for the case of a bipolar waveguide composed of a barrier potential, $u_{\mathrm{II}}(x)$, which is equal and opposite in sign to the well potential, $u_{\mathrm{I}}(x)$, i.e., $u_{\mathrm{II}}\left(x\right)=-u_{\mathrm{I}}\left(x-d\right)$, that it is sufficient to use the bound state solutions of the well alone to calculate the size of the pseudo-gaps. Let us consider the case of a single potential well, defined by the electrostatic potential $u(x)$, whose exact bound state solutions for graphene are known. The massless 2D Dirac equation of a barrier defined by the quasi-1D potential $-u(x)$ can be written as
\begin{equation}
\left[-i\sigma_{x}\frac{\partial}{\partial x}+k_{y}\sigma_{y}-\mathrm{I}u(x)\right]\Psi=E\Psi,
\nonumber
\end{equation}
which is formally equivalent to 
\begin{equation}
\left[-i\sigma_{x}\frac{\partial}{\partial x}+k_{y}\sigma_{y}+\mathrm{I}u(x)\right]\Psi^{\star}=-E\Psi^{\star}.
\nonumber
\end{equation}
Therefore, at $E=0$ the wavefunctions of the barrier are simply the complex conjugate of the identical well, and at non-zero energy, the barrier wavefunctions can be obtained from the well wavefunctions by taking its complex conjugate accompanied by an exchange of the sign of $E$. Let us now consider the particular case of a bipolar waveguide described by the potential given in Eq.~(\ref{eq:potential_s}), where
\begin{equation}
u_{\mathrm{I}}=-\frac{u_{0}}{\cosh\left[\left(x+\frac{d}{2}\right)/L\right]},\qquad u_{\mathrm{II}}=\frac{u_{0}}{\cosh\left[\left(x-\frac{d}{2}\right)/L\right]},
\label{eq:appendix_wells}
\end{equation}
and $u_0>0$. For the hyperbolic secant potential the bound state zero-energy solutions to Eq.~(\ref{eq:Ham_well}) (obtained in Ref. \cite{hartmann2010smooth}) for $k_y>0$ and $s_\mathrm{K}=1$ are
\begin{equation}
\Psi_{A}=i\left(-1\right)^{n+1}\,_{2}F_{1}\left(-n+2\omega,\,-n;\,1-n+\omega;\,t\right)t^{-\frac{n}{2}+\frac{\omega}{2}+\frac{1}{4}}\left(1-t\right)^{-\frac{n}{2}+\frac{\omega}{2}-\frac{1}{4}}
\label{eq:exact_A}
\end{equation}
and
\begin{equation}
\Psi_{B}=\,_{2}F_{1}\left(-n+2\omega,\,-n;\,1-n+\omega;\,1-t\right)t^{-\frac{n}{2}+\frac{\omega}{2}-\frac{1}{4}}\left(1-t\right)^{-\frac{n}{2}+\frac{\omega}{2}+\frac{1}{4}}
\label{eq:exact_B}
\end{equation}
where $\,_{2}F_{1}(a,b;c;x)$ is the hypergeometric function, $2t= 1-\tanh\left[\left(x+\frac{d}{2}\right)/L\right]$, $\omega=u_0/(\hbar v_{\mathrm{F}}$/L) and $n$ is a positive integer. It should be noted that for the same valley, the negative $k_y$ solutions can be obtained by simply exchanging $\Psi_{A}$ for $\Psi_{B}$ and vice-versa. Similarly for positive $k_y$, the opposite valley sub-lattice functions can be obtained from Eqs.~(\ref{eq:exact_A}-\ref{eq:exact_B}) via the interchange of $\Psi_{A}$ and $\Psi_{B}$. Substituting the exact solution for the $n=0$ mode into Eq.~(\ref{eq:energy_gap}) yields 
\begin{equation}
E_g=\frac{2^{1-2\omega}\hbar v_{\mathrm{F}}}{N_{\frac{d}{2}}N_{-\frac{d}{2}}}\left|\left(1-\rho\right)^{\omega+\frac{1}{2}}\left(1+\rho\right)^{\omega-\frac{1}{2}}+\left(1-\rho\right)^{\omega-\frac{1}{2}}\left(1+\rho\right)^{\omega+\frac{1}{2}}\right|,
\label{eq:energy_n0}
\end{equation}
where $\rho=\tanh\left(\frac{d}{2L}\right)$, and since $d\gg L$ Eq.~(\ref{eq:energy_n0}) becomes
\begin{equation}
E_{g}\approx\frac{2\hbar v_{\mathrm{F}}e^{-\left(\omega-\frac{1}{2}\right)\frac{d}{L}}}{LB\left(\omega+1/2,\omega-1/2\right)},
\nonumber
%E_g=\frac{2^{1-2\omega}\hbar v_{\mathrm{F}}}{N_{\frac{d}{2}}N_{-\frac{d}{2}}}\left(1-\rho\right)^{\omega-\frac{1}{2}}\left(1+\rho\right)^{\omega+\frac{1}{2}}.
\end{equation}
where $B(m,n)$ is the Beta function.

\section{Approximate expression for the transition matrix element}\label{Append_transition}
In the presence of an electromagnetic field, the particle momentum operator, $\hat{\boldsymbol{p}}$, is modified such that $\hat{\boldsymbol{p}}\rightarrow\hat{\boldsymbol{p}}+e\boldsymbol{A}/c$, where $e$ is the elementary charge, and $\boldsymbol{A}$ is the magnetic vector potential, which is related to $\textbf{e}=\left(e_{x},e_{y}\right)$, the unit vector describing the polarization of the electromagnetic wave, via the relation $\boldsymbol{A}=A\textbf{e}$. For linearly polarized light, the polarization vector is expressed as $\left(\cos\left(\varphi_{0}\right),\sin\left(\varphi_{0}\right)\right)
$, while for right- and left-handed polarized light it is
$\left(1,-i\right)/\sqrt{2}$ and $\left(1,i\right)/\sqrt{2}$, respectively. The general form of the perturbation due to an electromagnetic wave impinging normally to a Dirac material is
\begin{equation}
\delta H=
\frac{eA v_{\mathrm{F}}}{c}\left(\sigma_{x}e_{x}+s_{\mathrm{K}}\sigma_{y}e_{y}\right),
\nonumber
\end{equation}
which is related to the velocity operator, $\hat{\boldsymbol{v}}$, given by Eq.~(\ref{eq:VME}) in the main text, by the simple relation $\delta H=(eA/c)\hat{\boldsymbol{v}}\cdot\textbf{e}$~\cite{saroka2018momentum,hartmann2019interband}. The transition matrix element is proportional to $\left|\left\langle \Psi_{f}\left|\hat{\boldsymbol{v}}\cdot\textbf{e}\right|\Psi_{i}\right\rangle \right|^{2}$, where $\Psi_{i}$ and $\Psi_{f}$ are the initial and final states, respectively. Using the functions given in Eq.~(4) and Eq.~(5) of the main text, the transition matrix element at the pseudo-gap edge (or at the same value of momentum between the doublet states for two wells) in the small wavefunction overlap approximation is given by:
\begin{equation}
\left|\left\langle \Psi_{2}\left|\hat{\boldsymbol{v}}\cdot\textbf{e}\right|\Psi_{1}\right\rangle \right|/v_{\mathrm{F}}=\delta_{k_{y,i},k_{y,f}}\left|\intop_{-\infty}^{\infty}\left(\Phi_{0}+\Phi_{\pm}\right)dx\right|,
\label{eq:s_mat_opt}
\end{equation}
where the subscripts $+$ and $-$ correspond to the case of the double well and bipolar waveguide, respectively, $k_{y,i}$ and $k_{y,f}$ are the initial and final wavenumbers along the direction of the waveguide, and the terms $\Phi_{0}$ and $\Phi_{\pm}$ are defined as:
\begin{equation}
\begin{aligned}
\Phi_{0}&=i\,\mathrm{Im}\left[\Psi_{\pm}^{\dagger}\left(x-\frac{d}{2}\right)\left(\sigma_{x}e_{x}+s_{\mathrm{K}}\sigma_{y}e_{y}\right)\Psi_{0}\left(x+\frac{d}{2}\right)\right],\\
\Phi_{-}&=\Psi_{-}^{\dagger}\left(x-\frac{d}{2}\right)s_{\mathrm{K}}\sigma_{y}e_{y}\Psi_{-}\left(x-\frac{d}{2}\right),\\
\Phi_{+}&=0.\\
\end{aligned}
\nonumber
\end{equation}
Since the potentials $u_{\mathrm{I}}$ and $u_{\mathrm{II}}$ are symmetric,  the solutions of $\psi_{A}\left(x\pm\frac{d}{2}\right)$ and $\psi_{B}\left(x\pm\frac{d}{2}\right)$ can be constructed to be purely imaginary and real respectively. Therefore, $\Phi_{0}$,  which appears in the transition matrix element, Eq.~(\ref{eq:s_mat_opt}), reduces to \begin{equation}
\Phi_{0}=i\,\mathrm{Im}\left[\Psi_{\pm}^{\dagger}\left(x-\frac{d}{2}\right)\sigma_{x}e_{x}\Psi_{0}\left(x+\frac{d}{2}\right)\right].
\label{eq:s_mat_opt_temp}
\end{equation}
It should be noted that for pseudo-gaps arising from the repulsion of well and barrier zero-energy guided modes, $\Psi_{-}\left(x-\frac{d}{2}\right)=\Psi_{0}^{\star}\left(x-\frac{d}{2}\right)$, this coupled with the fact that $\psi_A$ is purely imaginary, and $\psi_B$, real, means that $\Phi_{0}$ appearing on the left hand side of Eq.~(\ref{eq:s_mat_opt_temp}) neither depends on the sign of $k_y$ nor on $s_{\mathrm{K}}$; whereas $\Phi_{-}$ is a function of $k_y$, i.e., $\Phi_{-}=\Phi_{-}(k_y)$, which obeys the relation $\Phi_{-}(-k_y)=-\Phi_{-}(k_y)$, and is independent of $s_{\mathrm{K}}$.

We will now determine the dipole matrix element for transitions across the $n=0$ pseudo-gap edge of a bipolar graphene waveguide, composed of a well and barrier of equal strength, described by the hyperbolic secant functions given in Eq.~(\ref{eq:appendix_wells}). In the limit that $d/L \gg 1$, $\Psi_{A}\left(x-\frac{d}{2}\right)\Psi_{B}\left(x+\frac{d}{2}\right) 
\gg \Psi_{A}\left(x+\frac{d}{2}\right)\Psi_{B}\left(x-\frac{d}{2}\right)$. Upon changing to the variable $\tilde{t}=\left\{ 1-\tanh\left[\left(x-\frac{d}{2}\right)/L\right]\right\} /2$, integrating $\Phi_{0}$ across the domain of $x$, and retaining only the $\Psi_{A}\left(x-\frac{d}{2}\right)\Psi_{B}\left(x+\frac{d}{2}\right)$ terms, one obtains
\begin{equation}
\intop_{-\infty}^{\infty}\Phi_{0}\,dx\approx-i\frac{e_{x}e^{-\Delta_{0}d/L}}{B\left(1+\Delta_{0},\,\Delta_{0}\right)}\intop_{0}^{1-e^{-d/L}}\tilde{t}^{-\frac{1}{2}+\Delta_{0}}\left(1-\tilde{t}\right)^{-1}d\tilde{t}.
\label{eq:bipolar_append_vme}
\end{equation}
The integral $\int_{0}^{1-e^{-d/L}}\tilde{t}^{-\frac{1}{2}+\Delta_{0}}\left(1-\tilde{t}\right)^{-1}d\tilde{t}$ is of the form of the incomplete Beta function, which we shall denote as $B(1-e^{-d/L};\frac{1}{2}+\Delta_{0},0)$. 
%It should be noted that repeating the same procedure for two equal strength hyperbolic secant potential wells yields the complex conjugate of Eq.~(\ref{eq:bipolar_append_vme}). 
The integration of $\Phi_-$ from $x=-\infty$ to $\infty$ yields
\begin{equation}
\intop_{-\infty}^{\infty}\Phi_{-}\,dx=-\frac{k_{y}}{\left|k_{y}\right|}\frac{e_{y}B\left(\frac{1}{2}+\Delta_{0},\frac{1}{2}+\Delta_{0}\right)}{B\left(1+\Delta_{0},\,\Delta_{0}\right)}.
\nonumber
\end{equation}
Therefore, for the bipolar waveguide, the transition matrix element at the $n=0$ pseudo-gap edge is
\begin{equation}
\left|\left\langle \Psi_{2}\left|\hat{\boldsymbol{v}}\cdot\textbf{e}\right|\Psi_{1}\right\rangle \right|/v_{\mathrm{F}}\approx\left|\frac{e_{x}B\left(1-e^{-d/L};\frac{1}{2}+\Delta_{0},\,0\right)e^{-\Delta_{0}d/L}}{B\left(1+\Delta_{0},\,\Delta_{0}\right)}-i\frac{k_{y}}{\left|k_{y}\right|}\frac{e_{y}B\left(\frac{1}{2}+\Delta_{0},\,\frac{1}{2}+\Delta_{0}\right)}{B\left(1+\Delta_{0},\,\Delta_{0}\right)}\right|,
%\left|\left\langle \Psi_{2}\left|\hat{\boldsymbol{v}}\right|\Psi_{1}\right\rangle /v_{\mathrm{F}}\right|\approx\left|\frac{e_{x}B\left(1-e^{-d/L};\frac{1}{2}+\Delta_{0},\,0\right)e^{-\Delta_{0}d/L}}{B\left(1+\Delta_{0},\,\Delta_{0}\right)}-is_{k_{y}}s_{\mathrm{K}}\frac{e_{y}B\left(\frac{1}{2}+\Delta_{0},\,\frac{1}{2}+\Delta_{0}\right)}{B\left(1+\Delta_{0},\,\Delta_{0}\right)}\right|,
\label{eq:s_bi_polar}
\end{equation}
while for the case of the double well, at the same value of momentum, the transition matrix element is 
\begin{equation}
\left|\left\langle \Psi_{2}\left|\hat{\boldsymbol{v}}\cdot\textbf{e}\right|\Psi_{1}\right\rangle \right|/v_{\mathrm{F}}\approx\left|\frac{e_{x}B\left(1-e^{-d/L};\frac{1}{2}+\Delta_{0},\,0\right)e^{-\Delta_{0}d/L}}{B\left(1+\Delta_{0},\,\Delta_{0}\right)}\right|.
\label{eq:s_well}
\end{equation}
The striking feature of Eq.~(\ref{eq:s_bi_polar}), which also appears in the main text as Eq.~10, is the apparent dependence of the pseudo-valley population on the handedness of the excitation. This is indeed allowed by symmetry, since the inversion symmetry is lifted in a bipolar waveguide by opposite top-gate potentials. This feature does not depend on the particular model chosen to describe the bipolar waveguide, as we show below. 

In the most general form the transition matrix element can be written as
\begin{equation}
\left|\left\langle f\left|
\hat{\boldsymbol{v}}\cdot\textbf{e}
\right|i\right\rangle \right|/v_{\mathrm{F}}=\left|\intop_{-\infty}^{\infty}\left[\Psi_{B,f}^{\star}\left(e_{x}+ie_{y}s_{\mathrm{K}}\right)\Psi_{A,i}+\Psi_{A,f}^{\star}\left(e_{x}-ie_{y}s_{\mathrm{K}}\right)\Psi_{B,i}\right]dx\right|,
\label{eq:full_VME_s}
\end{equation}
where the subscripts $i$ and $f$ correspond to the initial and final states. The spinor components $\Psi_{A}$ and $\Psi_{B}$ depend on both the valley index number, $s_{\mathrm{K}}$, and the sign of $k_y$, i.e., $\Psi_{B,f}=\Psi_{B,f}(s_{\mathrm{K}},s)$, where $s=k_y/\left|k_{y}\right|$. From Eq.~(2) of the main text it is clear that the change of the sign of $k_y$, as well as $s_{\mathrm{K}}$, lead to the swapping of the $A$ and $B$ indices in Eq.~(\ref{eq:full_VME_s}). For $s=1$ and $s_{\mathrm{K}}=1$, Eq.~(\ref{eq:full_VME_s}) yields
\begin{equation}
\left|\left\langle f\left|\hat{\boldsymbol{v}}\cdot\textbf{e}
\right|i\right\rangle \right|/v_{\mathrm{F}}
=\left|\intop_{-\infty}^{\infty}\left[\Psi_{B,f}^{\star}\left(e_{x}+ie_{y}\right)\Psi_{A,i}+\Psi_{A,f}^{\star}\left(e_{x}-ie_{y}\right)\Psi_{B,i}\right]dx\right|. 
\label{eq:VME_1}
\end{equation}
For $s=1$ and $s_{\mathrm{K}}=-1$,
\begin{equation}
\left|\left\langle f\left|\hat{\boldsymbol{v}}\cdot\textbf{e}\right|i\right\rangle \right|/v_{\mathrm{F}}
=\left|\intop_{-\infty}^{\infty}\left[\Psi_{B,f}^{\star}\left(e_{x}+ie_{y}\right)\Psi_{A,i}+\Psi_{A,f}^{\star}\left(e_{x}-ie_{y}\right)\Psi_{B,i}\right]dx\right|.
\label{eq:VME_2}
\end{equation}
For $s=-1$ and $s_{\mathrm{K}}=1$, 
\begin{equation}
\left|\left\langle f\left|\hat{\boldsymbol{v}}\cdot\textbf{e}\right|i\right\rangle \right|/v_{\mathrm{F}}=\left|\intop_{-\infty}^{\infty}\left[\Psi_{B,f}^{\star}\left(e_{x}-ie_{y}\right)\Psi_{A,i}+\Psi_{A,f}^{\star}\left(e_{x}+ie_{y}\right)\Psi_{B,i}\right]dx\right|. 
\label{eq:VME_3}
\end{equation}
For $s=-1$ and $s_{\mathrm{K}}=-1$,
\begin{equation}
\left|\left\langle f\left|\hat{\boldsymbol{v}}\cdot\textbf{e}\right|i\right\rangle \right|/v_{\mathrm{F}}=\left|\intop_{-\infty}^{\infty}\left[\Psi_{B,f}^{\star}\left(e_{x}-ie_{y}\right)\Psi_{A,i}+\Psi_{A,f}^{\star}\left(e_{x}+ie_{y}\right)\Psi_{B,i}\right]dx\right|.
\label{eq:VME_4}
\end{equation}
combining Eqs~(\ref{eq:VME_1}-\ref{eq:VME_4}) together results in an expression which does not depend on the sign of $s_\mathrm{K}$:
\begin{equation}
\left|\left\langle f\left|\hat{\boldsymbol{v}}\cdot\textbf{e}\right|i\right\rangle /v_{\mathrm{F}}\right|=\left|\intop_{-\infty}^{\infty}\left[\Psi_{B,f}^{\star}\left(e_{x}+ise_{y}\right)\Psi_{A,i}+\Psi_{A,f}^{\star}\left(e_{x}-ise_{y}\right)\Psi_{B,i}\right]dx\right|,
\label{eq:VME_5}
\end{equation}
where the functions $\Psi_A$ and $\Psi_B$ entering Eqs~(\ref{eq:VME_1}-\ref{eq:VME_5}) are evaluated for $s_\mathrm{K}=s=1$.

\bibliography{ref}% Produces the bibliography via BibTeX.

\end{document}